\documentclass[floatfix, aps,prb,twocolumn, groupedaddress]{revtex4-2}
\usepackage{dcolumn}
\usepackage{bm}
\usepackage{amssymb}
\usepackage{amsthm}
\usepackage{physics}
\usepackage{color}
\usepackage[version=4]{mhchem}
\usepackage{graphicx}
\usepackage{natbib}
\usepackage{subfig}
\usepackage{booktabs}
\usepackage{multirow}
\usepackage[T1]{fontenc}
\newcommand{\degC}{$^\circ$C}
\newcommand{\etal}{\textit{et al}.}

\newcommand{\Biso}{B_{\rm iso}}
\newcommand{\TC}{T_{\rm C}}

\newcommand{\TCC}{T^*}
\newcommand{\TCW}{\theta_{\rm W}}
\newcommand{\peff}{p_{\rm eff}}
\newcommand{\uB}{\mu_{\rm B}}
\newcommand{\kB}{k_{\rm B}}
\newcommand{\Msat}{M_{\rm sat}}
\newcommand{\Mspon}{M_{\rm s}}

\newcommand{\Hm}{H_{\rm m}}

\newcommand{\dSM}{\Delta S_{\rm M}}
\newcommand{\dSMax}{\Delta S_{\rm M, max}}
\newcommand{\invchi}{\chi_0^{-1}}
\newcommand{\EF}{E_{\rm F}}
\newcommand{\FMAo}{FM$_{\rm aniso}$1}
\newcommand{\FMAt}{FM$_{\rm aniso}$2}
\newcommand{\LIC}{\ce{LuInCo_4}}
\newcommand{\LC}{\ce{LuCo_2}}
\newcommand{\YMC}{\ce{YMgCo_4}}
\newcommand{\RIC}{\ce{$R$InCo_4}}
\newcommand{\RMC}{\ce{$R$MgCo_4}}
\newcommand{\RC}{\ce{$R$Co_2}}
\newcommand{\YC}{\ce{YCo_2}}
\newcommand{\LuC}{\ce{LuCo_2}}

\newcommand{\LF}{\ce{LuFe_2}}

\begin{document}


\title{Unusual strong ferromagnetism in site-ordered cubic Laves phase compound {\LIC} with Co-pyrochlore lattice}


\author{Taiki Shiotani}
\email[]{Corresponding author: shiotani.taiki.48e@st.kyoto-u.ac.jp}
\affiliation{Department of Materials Science and Engineering, Kyoto University, Kyoto 606-8501, Japan}

\author{Hiroto Ohta}
\affiliation{Department of Molecular Chemistry and Biochemistry, Doshisha University, Kyotanabe 610-0321, Japan}

\author{Takeshi Waki}
\affiliation{Department of Materials Science and Engineering, Kyoto University, Kyoto 606-8501, Japan}

\author{Yoshikazu Tabata}
\affiliation{Department of Materials Science and Engineering, Kyoto University, Kyoto 606-8501, Japan}

\author{Hiroyuki Nakamura}
\affiliation{Department of Materials Science and Engineering, Kyoto University, Kyoto 606-8501, Japan}

\date{\today}

\begin{abstract}
We have successfully synthesized single crystals of the site-ordered cubic (C15b) Laves phase compound {\LIC} with the Co-pyrochlore sublattice for the first time. {\LIC} undergoes a ferromagnetic transition at 306 K and has a saturation moment of 3.43\ $\uB$/{f.u.}\ at 5\ K. The strongly ferromagnetic nature was verified by DFT calculations, suggesting that Co--3$d$ flat bands near the Fermi level induce the spin polarization. The magnetization is isotropic above $\sim$100\ K, saturates most easily in the [100] direction at low temperatures. In this anisotropic ferromagnetic state, the magnetization undergoes a metamagnetic transition in the [111] direction. Our results suggest that {\LIC} is a new strong but unusual itinerant electron ferromagnet, which deserves further study as a pyrochlore metal.
\end{abstract}


\maketitle

\section{Introduction}
\begin{figure}[h]
  \centering
  \includegraphics[keepaspectratio, width=0.7\columnwidth]{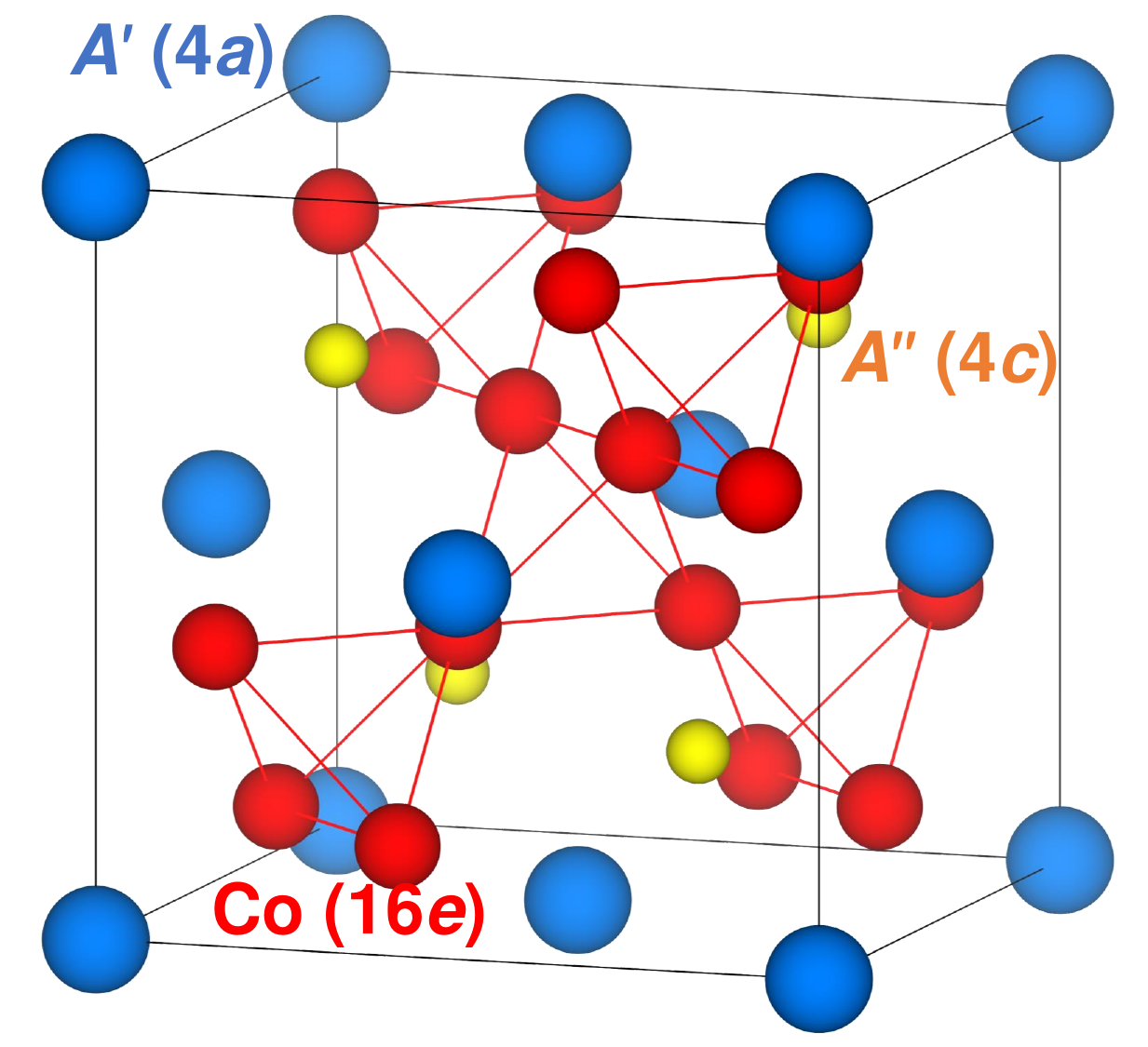}
  \caption{Crystal structure of the C15b-type Laves phase compound. The $A^{\rm \prime}$ site is occupied by a rare earth element, and $A^{\rm \prime \prime}$ by Mg or In.}
  \label{cryst}
\end{figure}
\indent The interplay between unique lattice geometry and electronic correlations can lead to intriguing physical properties. The pyrochlore lattice, a network of corner-sharing tetrahedra, is a prototype of geometrically frustrated systems. 
Novel magnetic states due to frustration on the pyrochlore lattice have been discovered in a variety of insulators. Representative phenomena of the pyrochlore lattice are a spin ice in the rare-earth titanium oxides {\ce{Dy_2Ti_2O_7}} and {\ce{Ho_2Ti_2O_7}} \cite{spiniceDy,spiniceHo}, spin glass in stoichiometrically pure {\ce{Y_2Mo_2O_7}} \cite{spigla}, and collective paramagnetism or spin liquid behavior in {\ce{Tb_2Ti_2O_7}} \cite{QSL}. 
Although less explored than for insulators, the role of the pyrochlore lattice on physical properties in the itinerant electron system is also being understood. For example, the C15 cubic Laves phase {\ce{YMn_2}} exhibits helical magnetic ordering with an extremely long period \cite{YMheli} and a spin-liquid state with small Sc substitution due to frustration on the Mn pyrochlore lattice \cite{QSLYMn}.
Recently, non-trivial band structures in pyrochlore metals have also attracted attention. Analogous to the kagome lattice, a two-dimensional network of corner-sharing triangles, pyrochlore metals have been proposed to host three-dimensional flat bands due to quantum destructive interference \cite{flata, flatb}, which have been experimentally discovered in the C15 cubic Laves phase compound {\ce{CaNi_2}} and the thiospinel compound {\ce{CuV_2S_4}} \cite{CaNi,CuVS}. In addition, theoretical studies have predicted a four-fold degenerate Dirac point at the high symmetry point of the Brillouin zone \cite{diraca, diracb, diracc}.
\\
\indent A family of intermetallics with a Co pyrochlore lattice is known as the C15 Laves phase {\RC} ($R$: rare-earth element). In {\RC}, despite the large density of states (DOS) of the Co--3$d$ bands just below the Fermi level, the Co sublattice has no spontaneous magnetic moment, but the Co--3$d$ bands are magnetically polarized by the molecular field from the magnetic $R$ sublattice \cite{RC}. Therefore, {\YC} and {\LuC} are exchange-enhanced Pauli paramagnets with high magnetic susceptibility and electronic specific heat coefficients \cite{Ymeta, Lumeta, lemaire, burzo}.
Recently, a new series of cubic Laves phase {\RMC} ($R$\ =\ Y, Ce, Gd--Tm and Lu) with an $A$-site ordered (C15b) structure [Fig.\ \ref{cryst}] has been reported on synthesis and magnetism \cite{QQYMC, VVYMC, YMC, CeMC, TbMC, pottgen}. In $\RMC$, Co atoms occupy the Wyckoff position 16$e$ ($x$, $x$, $x$), $R$ atoms 4$a$ (0, 0, 0), and Mg atoms 4$c$ ($\frac{1}{4}$, $\frac{1}{4}$, $\frac{1}{4}$). Interestingly, {\RMC} has two frustrated sublattices: a breathing but nearly ideal pyrochlore sublattice formed by Co atoms and a face-centered cubic sublattice formed by $R$ atoms, possibly leading to versatile quantum phases by changing the $R$ atom.   
So far, {\YMC} has been reported to show strong ferromagnetism with a Curie temperature $\TC = 405$\ K \cite{YMC}, in contrast to {\YC}, while {\ce{TbMgCo_4}} has a ferrimagnetic structure where the Co moments are antiparallel to the Tb moments \cite{TbMC}.
It is noteworthy that $^{59}$Co-NMR measurements revealed that {\YMC} has widely distributed internal field at low temperature. This indicates that {\YMC} may not have a simple collinear structure, but rather a helically modulated ferromagnetic structure \cite{YMC}.
Although these features provide a new platform for exploring nontrivial magnetic phenomena in pyrochlore metals, it is limited to investigate the detailed physical properties. This is because the high vapor pressure of Mg makes it difficult to grow single crystals of {\RMC}.
\\
\indent In this context, we focused on {\RIC} ($R$\ =\ Dy--Lu), which is isostructural with {\RMC}. Its presence and crystal structure have been reported by Sysa {\etal} in 1988 \cite{LIC}. Because the 4$c$ site is occupied by In instead of Mg, it is expected to be easier to synthesize single crystals. However, to date, there have been no reports on the growth and physical properties of the single crystals. 
In this paper, we report the single crystal synthesis of {\LIC} and discuss its magnetic properties based on magnetization measurements and density functional theory (DFT) calculations. We newly found that {\LIC} exhibits a ferromagnetic transition at $\TC = 306$\ K with a saturation moment of 3.43\ $\uB/${f.u.}\ at $T=5$\ K. We also report a metamagnetic transition at low temperature under the field applied in the [111] direction, suggesting that {\LIC} does not belong to the typical ferromagnets.\\
\section{Experiments}
We have synthesized {\LIC} single crystals by the self-flux method starting from Lu ingots (Nippon Yttrium, 99.9\% purity), In shots (Rare Metallic, 99.99\%), and Co flakes (Rare Metallic, 99.9\%).
The mixture with a composition of {\ce{LuIn_2Co_2}} was arc-melted in an argon atmosphere, followed by grinding. The resulting mixture was placed in a BN crucible and sealed in an evacuated quartz tube. The ampoule was initially heated up to 1200\ \degC\ over 3 h and held for 0.5 h, then cooled down to 1150 \degC\ over 8 h, and slowly cooled down to 975\ \degC\ over 58h, at which the ampoule was centrifuged to separate single crystals from the flux. Octahedral or triangular crystals with well-developed \{111\} faces were obtained.
The flux and impurity phases remaining on the surface were removed by immersion in dilute HClaq.
\\
\indent The grown crystals were characterized by x-ray diffraction (XRD) measurements with Cu\ $K_{\alpha_1}$ radiation using X’Pert PRO Alpha-1 (PANalytical). The Rietveld refinement was performed using Rietan-FP \cite{FP}. For this purpose powder was prepared by crushing a single crystal.
Magnetization measurements were performed using a SQUID magnetometer (MPMS, Quantum Design) in the temperature range of 5--350 K and under magnetic fields up to 7 T. We used octahedral crystals.\\
\indent DFT calculations were performed using the Vienna \textit{ab} initio simulation package (VASP) \cite{VASPa, VASPb, VASPc, VASPd}. We used the projector augmented wave (PAW) pseudopotentials \cite{PAWa,PAWb} with the generalized gradient approximation (GGA) scheme following the Perdew, Burke and Ernzerhof (PBE) functional \cite{GGA}. The conjugate gradient algorithm \cite{CGA} was used for structural relaxation. The tetrahedron method with Bl\"{o}chl corrections \cite{Blochl} and the Methfessel-Paxton scheme \cite{MP} were used for both geometry relaxation and total energy calculations. All atoms were relaxed until the forces on the atoms were less than $10^{-2}$ eV/\AA\ and the energy difference between two successive electronic steps was less than $10^{-7}$ eV. The $k$-point mesh along the high symmetry path was set to 50.\\
\section{Results}
\subsection{Crystal structure}
\begin{figure}[ht]
    \centering
    \includegraphics[keepaspectratio, width=1\columnwidth]{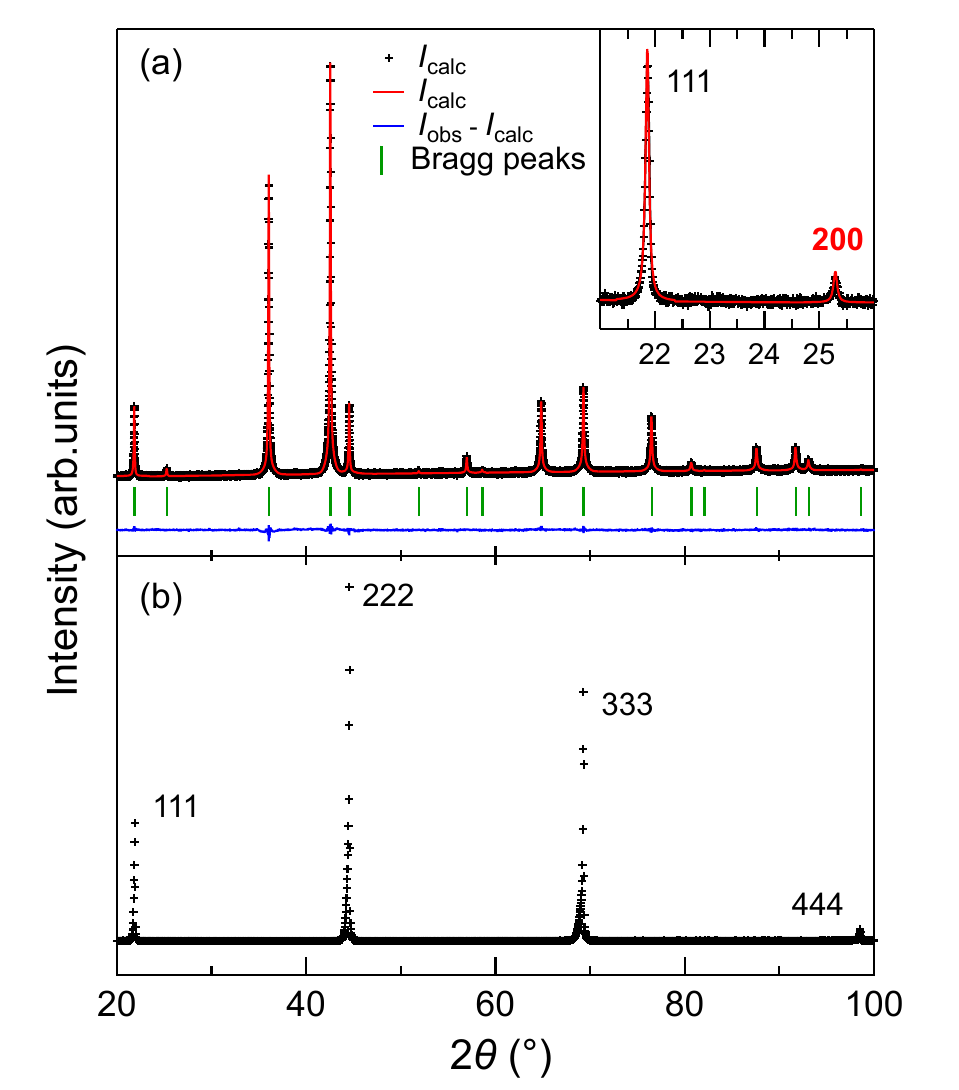}
    \caption{(a) Powder XRD profile of {\LIC} at room temperature. The result of the Rietveld refinement and expected Bragg reflections are also shown. The inset shows the data in the extended $2\theta$ range where the 200 superstructure peak is observed. (b) Monochromatic XRD pattern of the {111}-surface of the {\LIC} single crystal.}
    \label{xrd}
\end{figure}
Figure\ \ref{xrd}(a) shows the powder XRD profiles of {\LIC} at room temperature. We successfully obtained a single phase of {\LIC}. The 200 superstructure peak characteristic of the C15b-type structure was observed at $2\theta \simeq 25^{\circ}$ (see the inset of Fig.\ \ref{xrd}(a)), indicating that the reflection can be indexed by the planes of the site-ordered cubic Laves phase with space group $F\bar{4}3m$. 
The monochromatic XRD pattern collected from the flat surface of the single crystal showed the as-grown facet with the $\langle 111 \rangle$ direction perpendicular to the plate [Fig.\ \ref{xrd}(b)]. 
We performed structure refinement by the Rietveld method using Rietan-FP. 
Since optimizing all parameters at once resulted in negative isotropic displacement parameters $\Biso$, which are physically unreasonable, all $\Biso$ were fixed as follows: $\Biso({\rm Lu})=\Biso({\rm In})=0.75$ and $\Biso({\rm Co})=0.7$, with reference to literature data \cite{YbInCu, pottgen}. 
The lattice parameter was estimated to be $a = 7.03874(1)$\ {\AA}, in agreement with literature data (7.029 {\AA} \cite{LIC}). 
The atomic coordinate at the Co\ $16e\ (x, x, x)$ site was calculated to be $x=0.6252(2)$, which is close to that in the non-breathing pyrochlore lattice ($x=0.625$). The nearest and next nearest neighbor intraatomic distances in the breathing pyrochlore lattice, 2.485(2) and 2.492(2) \AA, respectively, are shorter than 2.53675 {\AA} in the non-breathing pyrochlore lattice of {\LC} \cite{LC}.
The site occupancy of Lu/In at the 4$a$ site and In/Lu at the 4$c$ site were 0.99(6)/0.01(7) and 0.87(6)/0.08(4), respectively, verifying the almost perfect atomic ordering of Lu and In in {\LIC}. 
\\
\subsection{Magnetization}
\begin{figure}[ht]
    \centering
    \includegraphics[keepaspectratio, width=1\columnwidth]{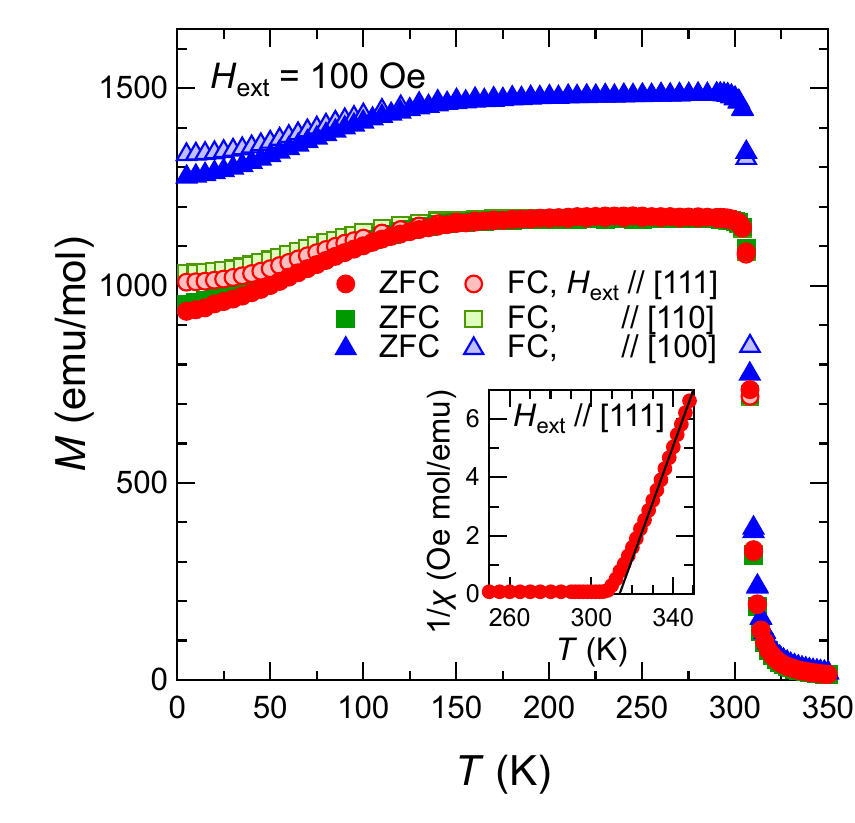}
    \caption{Temperature dependence of the magnetization of {\LIC} at $H_{\rm ext}=100$\ Oe. Open and filled markers represent the data under field-cooled and zero-field-cooled conditions, respectively. The inset shows the inverse susceptibility. The line represents the fit to the Curie-Weiss law. }
    \label{MT}
\end{figure}
\begin{figure*}[ht]
    \begin{tabular}{cccc}
      \begin{minipage}[h]{0.32\linewidth}
        \centering
        \includegraphics[keepaspectratio, width=\columnwidth]{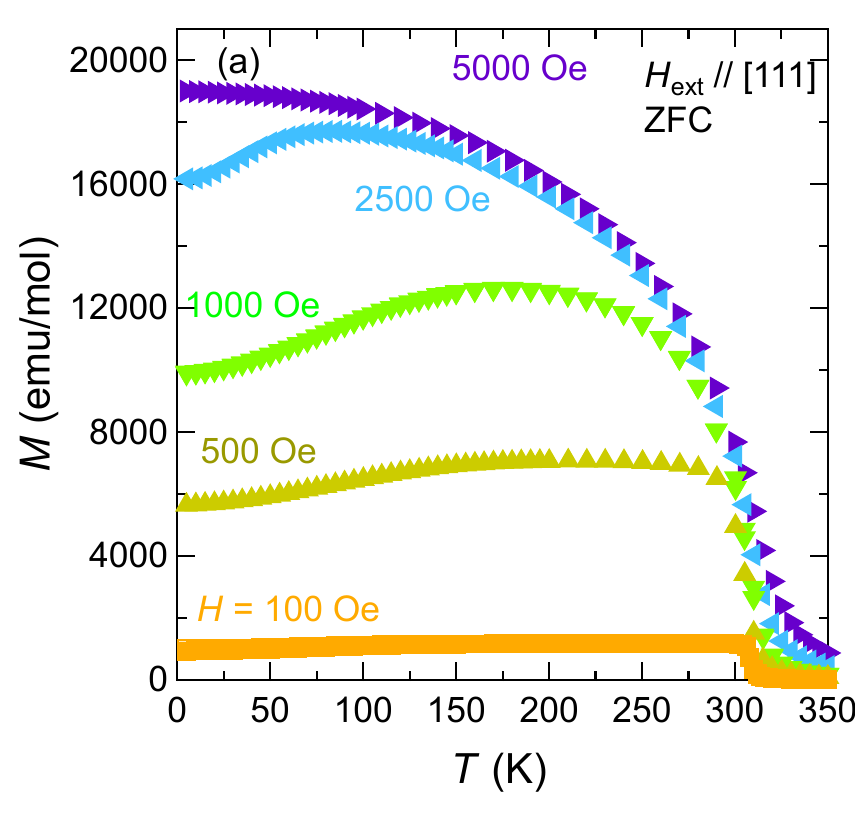}
      \end{minipage}&
      \begin{minipage}[h]{0.13\linewidth}
        \centering
        \includegraphics[keepaspectratio, width=\columnwidth]{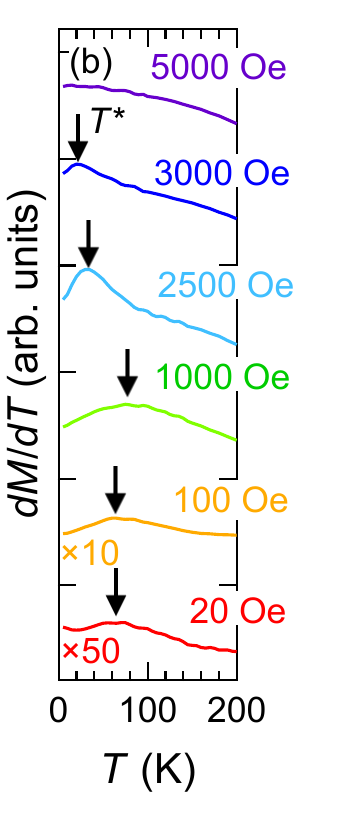}
      \end{minipage}&
      \begin{minipage}[h]{0.32\linewidth}
        \centering
        \includegraphics[keepaspectratio, width=\columnwidth]{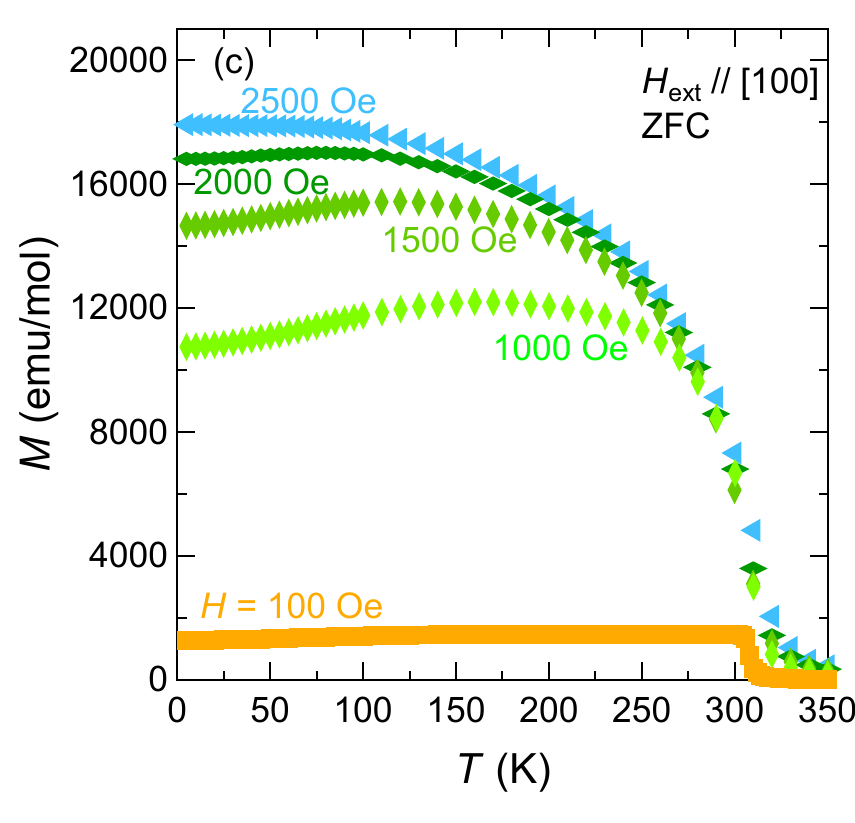}
      \end{minipage}&
      \begin{minipage}[h]{0.13\linewidth}
        \centering
        \includegraphics[keepaspectratio, width=\columnwidth]{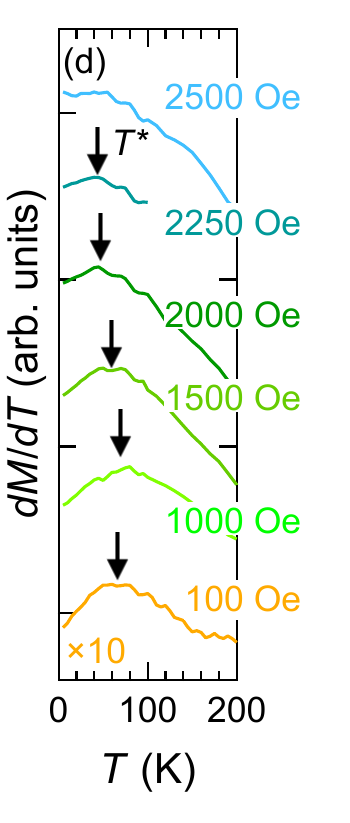}
      \end{minipage}
    \end{tabular}
      \caption{(a)(c) Temperature dependence of the magnetization and (b)(d) its derivative $dM/dT$ under magnetic fields along the [111] and [100] axes.}
      \label{mtdmdt}
    \end{figure*}
Figure\ \ref{MT} shows the temperature dependence of the magnetization of {\LIC} in the external field of $H_{\rm ext}=100$\ Oe. It shows a ferromagnetic transition at a Curie temperature of $\TC \simeq 310$\ K. 
The susceptibility above $\TC$ was well fitted by the Curie-Weiss law $\chi = C/(T-\TCW)$, where $C$ is the Curie constant and $\TCW$ is the Weiss temperature. The inverse susceptibility $1/\chi$ under the field applied in the [111] direction and its fitting result are shown in the inset of Fig.\ \ref{MT}. The estimated value of $\TCW = 313.8$\ K is close to $\TC$, and suggests dominant ferromagnetic correlations in {\LIC}. The effective moment $\peff$, estimated using the relation $C=\uB{\peff}^2/3\kB$, is 3.22\ $\uB/$Co. These values do not depend significantly on the applied field direction.
The magnetization begins to decrease below $T\simeq 150$\ K, accompanied by a small bifurcation between zero-field-cooled and field-cooled data at $H_{\rm ext} = 100$\ Oe [Fig.\ \ref{MT}]. The bifurcation becomes weaker with increasing field and almost disappears at $H_{\rm ext}\simeq 1$\ kOe. 
Figure\ \ref{mtdmdt} shows the temperature-dependent magnetization at different fields and its field derivatives. We found a peak in $dM/dT$ at the low temperature. The corresponding temperature $\TCC$ and the temperature range where the magnetization is reduced change with the applied field strength. At $H_{\rm ext} \simeq 1$\ kOe in the [111] direction, $\TCC$ decreases. At $H_{\rm ext} \simeq 5$\ kOe, it reaches 0 K, resulting in a monotonic increase of the magnetization. 
We observed the same behavior for the field applied in the [110] direction. When $H_{\rm ext} \parallel [100]$, the anomaly disappears at a lower field of $H_{\rm ext} \simeq 2.5$\ kOe [Fig.\ \ref{mtdmdt}(b)].
\\
\subsubsection{Anisotropic magnetic isotherms}
\begin{figure*}
  \begin{minipage}[h]{0.45\linewidth}
      \centering
      \includegraphics[keepaspectratio, width=1\columnwidth]{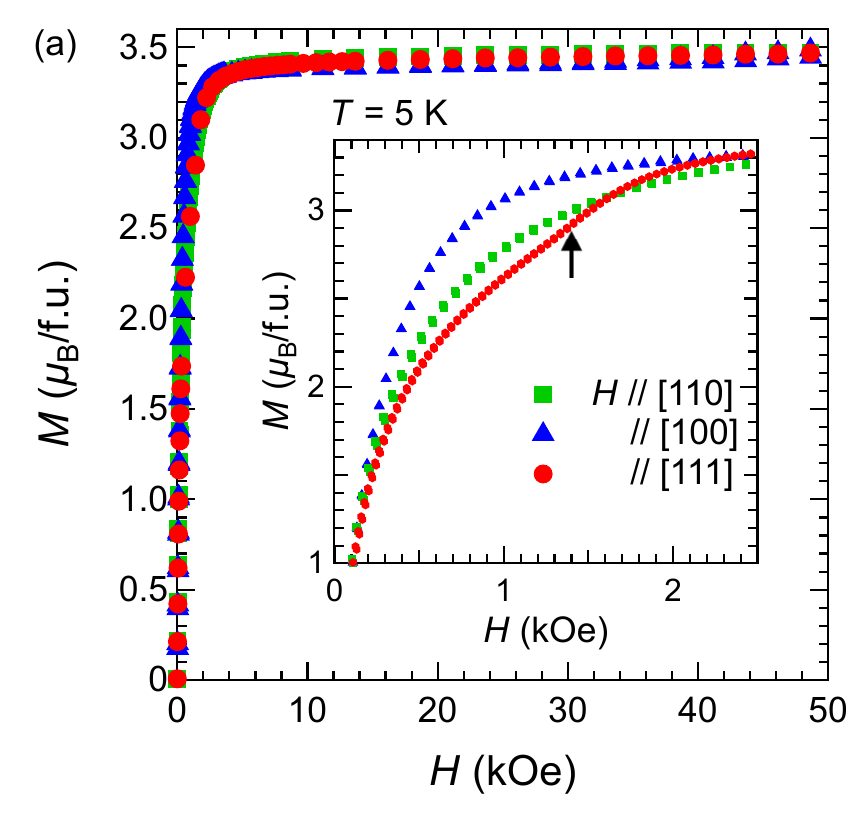}
    \end{minipage}
    \begin{minipage}[h]{0.45\linewidth} 
    \begin{tabular}{cc}
        \begin{minipage}[h]{0.48\linewidth}
          \centering
          \includegraphics[keepaspectratio, width=\columnwidth]{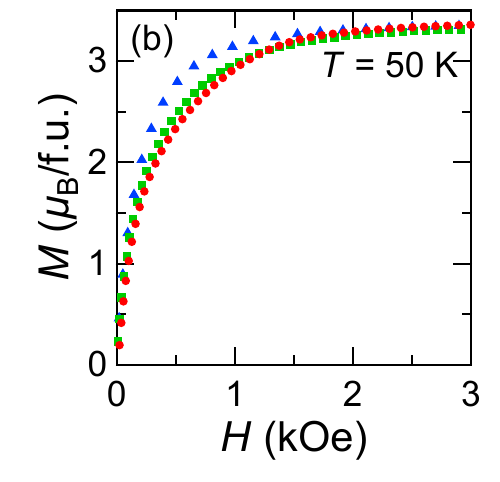}
        \end{minipage}&
        \begin{minipage}[h]{0.48\linewidth}
          \centering
          \includegraphics[keepaspectratio, width=\columnwidth]{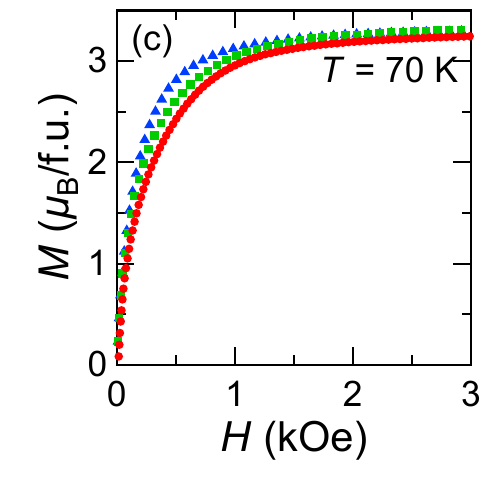}
        \end{minipage}\\
        \begin{minipage}[h]{0.48\linewidth}
          \centering
          \includegraphics[keepaspectratio, width=\columnwidth]{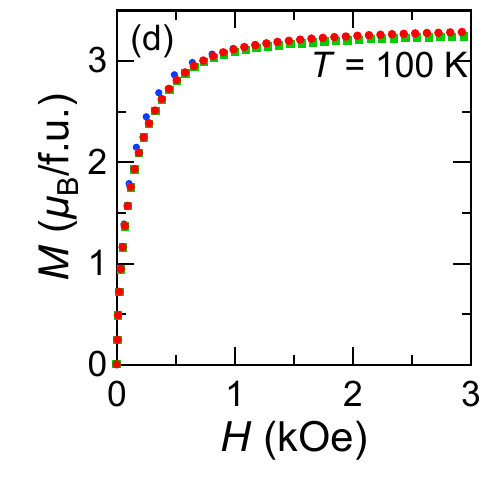}
        \end{minipage}&
        \begin{minipage}[h]{0.48\linewidth}
          \centering
          \includegraphics[keepaspectratio, width=\columnwidth]{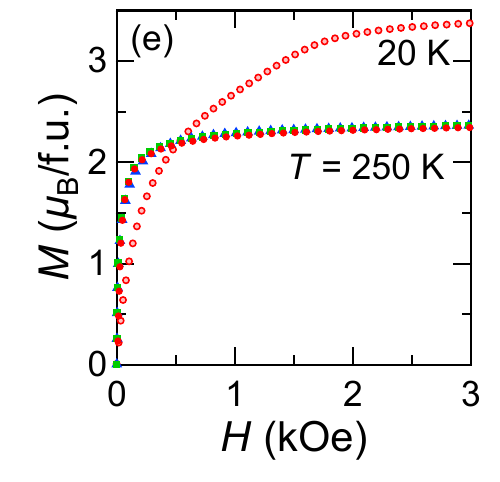}
        \end{minipage}
      \end{tabular}
    \end{minipage}
        \caption{Magnetization curves of {\LIC} at (a) $T=5$\ K, (b) 50\ K, (c) 70\ K, (d) 100\ K, and (e) 250\ K for the single crystal. The inset of Fig. \ref{mhmh}(a) shows the data in the low-field region. The case of $T = 20$\ K and $H\parallel [111]$ is also included in Fig. \ref{mhmh}(e).}
        \label{mhmh}
  \end{figure*}
  \begin{figure}[ht]
    \centering
    \includegraphics[keepaspectratio, width=1\columnwidth]{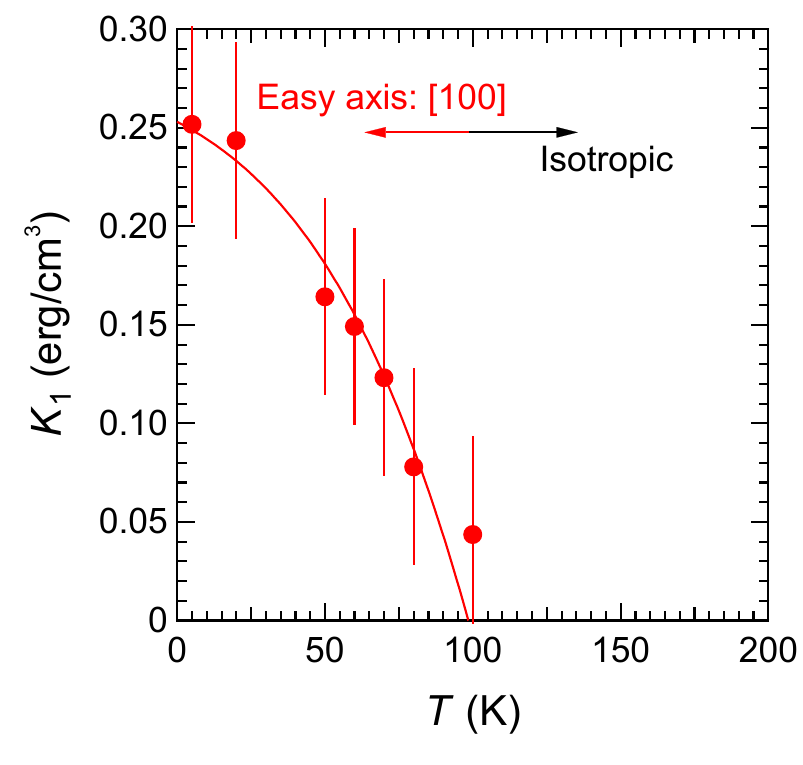}
    \caption{Temperature dependence of the magnetocrystalline anisotropy constant $K_1$ of {\LIC}.}
    \label{aniso}
  \end{figure}
\indent To reveal the magnetic anisotropy in {\LIC}, we measured the field-dependent magnetization at different temperatures. The crystal in the form of a regular octahedron was used, and the magnetic field was corrected by subtracting the demagnetizing field as $H = H_{\rm ext} - N M$, where $N$ is the demagnetization factor. $N$ for the [100], [110], and [111] directions were calculated to be $N \simeq 0.31, 0.29$, and 0.29, respectively, from the inverse slope of the low-field magnetization curve ($M(H_{\rm ext})/H_{\rm ext} \approx 1/D$) in the ferromagnetic state just below $\TC$. These nearly isotropic values close to $N=1/3$ reflect that the regular octahedron has a geometry close to a sphere, leading to the validity of the estimate.
Figure\ \ref{mhmh}(a) shows the field dependence of the magnetization of the single crystal at $T=5$\ K. It shows ferromagnetic behavior with a saturation magnetization of $\Msat \simeq 3.43\ \uB/$f.u. The inset of Fig.\ \ref{mhmh}(a) shows that the easy magnetization axis is in the [100] direction. As can be seen from Fig.\ \ref{mhmh}, the difference in the magnetization between each crystallographic axis gradually becomes less pronounced with increasing temperature, and the magnetization process becomes isotropic above $T = 100$\ K. 
The slope of the magnetization in the low field is larger at the higher temperatures (see Fig.\ \ref{mhmh}(e)). This means that the magnetization saturates more easily in the high temperature region, which is in accordance with the small reduction in the temperature-dependent magnetization at low temperature.\\
\indent The anisotropy energy of a cubic ferromagnet can be written simply as
\begin{equation}
  E_{\rm A} = K_0 + K_1(a_1^2 a_2^2 + a_2^2 a_3^2 + a_3^2 a_1^2)+K_2 a_1^2 a_2^2 a_3^2,
\end{equation}
where $K_0$, $K_1$ and $K_2$ are the anisotropy constants and $a_1$, $a_2$, $a_3$ are the directional cosines of the magnetization with respect to the cubic axes. $K_1$ and $K_2$ can be estimated thermodynamically. The energy required to magnetize a crystal to saturation ($M=\Msat$) in a direction $[hkl]$ is given by $E_{hkl}=\int_{0}^{\Msat}H_{hkl}dM_{hkl}$.
Comparing $E_{\rm A}$ and $E_{hkl}$, $K_1$ and $K_2$ can be expressed as follows:
\begin{align}
  &K_1 = 4(E_{110}-E_{100}),\\
& K_2 = 9(E_{100}+3E_{111}-4E_{110}).
\end{align}
Evaluating the area of the magnetization process up to saturation [Fig.\ \ref{mhmh}(a)] yields $K_1=0.25$ and $K_2=0.13$\ erg/cm$^3$ ($K_1=0.025$ and $K_2=0.013$\ MJ/m$^3$) at $T=5$\ K. $K_1$ is of the same order of magnitude as iron metal ($K_1 \simeq 0.48$\ erg/cm$^3$ \cite{Blundell}). 
Figure\ \ref{aniso} shows the temperature dependence of $K_1$, which decreases monotonically with the increasing temperature and reaches 0 at $T \simeq 100$\ K. This dependence demonstrates that the magnetic anisotropy changes from the easy magnetization axis along the [100] direction to isotropic as shown in Fig.\ \ref{mhmh}. 
\\
\subsubsection{Metamagnetic transition at low temperature}
\begin{figure}[ht]
  \centering
  \includegraphics[keepaspectratio, width=1\columnwidth]{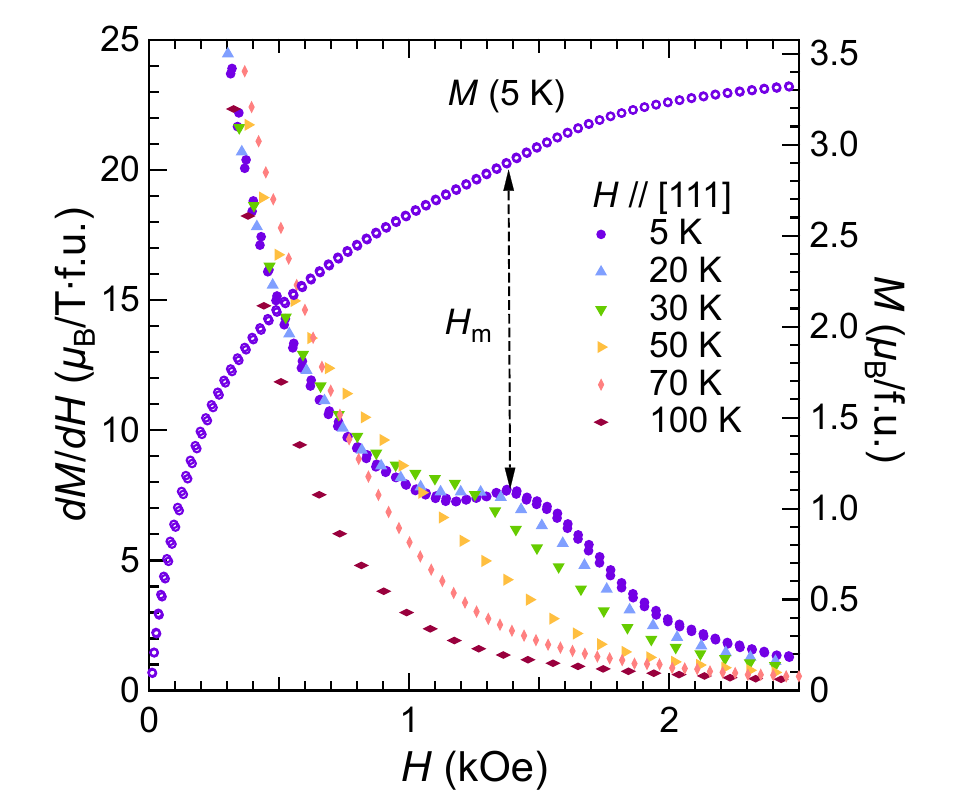}
  \caption{Field derivative of the magnetization curve of {\LIC} with the field applied in the [111] direction. Open circles show the magnetization curve at $T=5$\ K.}
  \label{dmdh}
\end{figure}
\indent From inset of the Fig.\ \ref{MT} we can see a small increase in the magnetization at $H\simeq 1.4$\ kOe only in the [111] direction. This anomaly can be clearly observed as a peak in its field derivative [Fig.\ \ref{dmdh}]. 
The peak of $dM/dH$ changes to a shoulder shape with raising temperature, and the metamagnetic-like bulge changes to a kink. The anomaly of $dM/dH$ shifts to lower fields and disappears at $T\simeq 100$\ K, where the magnetization becomes isotropic [Fig.\ \ref{aniso}]. This feature is different from that expected for a standard ferromagnet, suggesting the presence of different magnetic phases.
\\
\subsubsection{Critical scaling analysis of magnetization}
\begin{figure*}[ht]
  \begin{tabular}{cc}
    \begin{minipage}[h]{0.48\linewidth}
      \centering
      \includegraphics[keepaspectratio, width=\columnwidth]{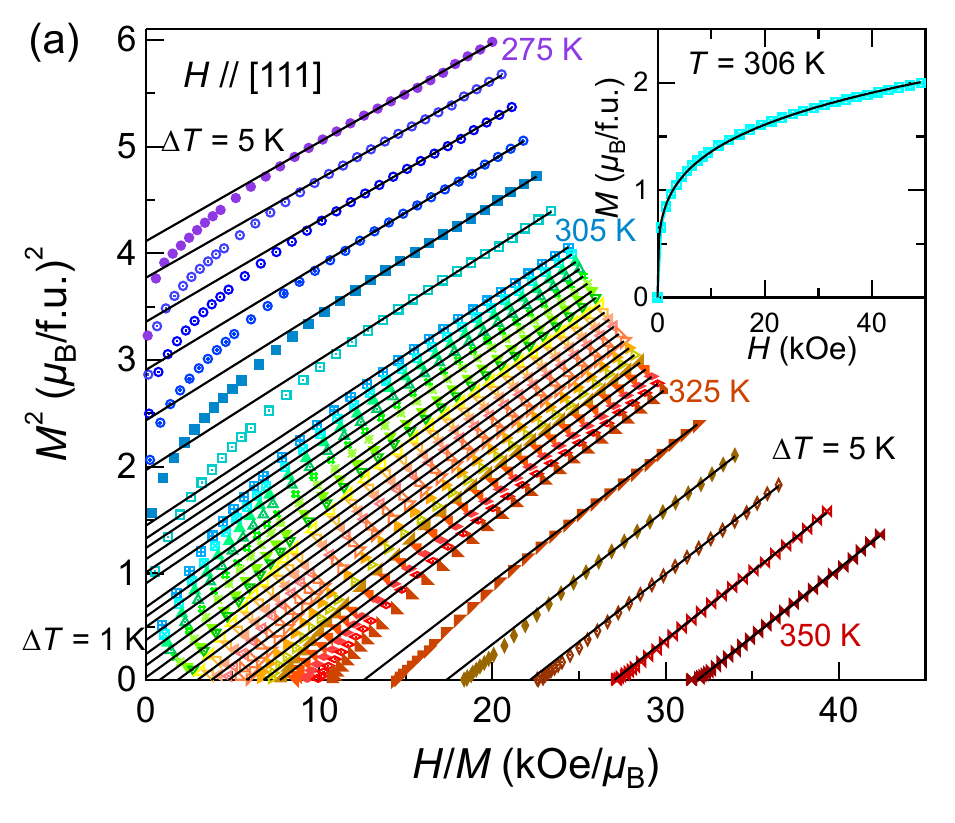}
    \end{minipage}&
    \begin{minipage}[h]{0.48\linewidth}
      \centering
      \includegraphics[keepaspectratio, width=\columnwidth]{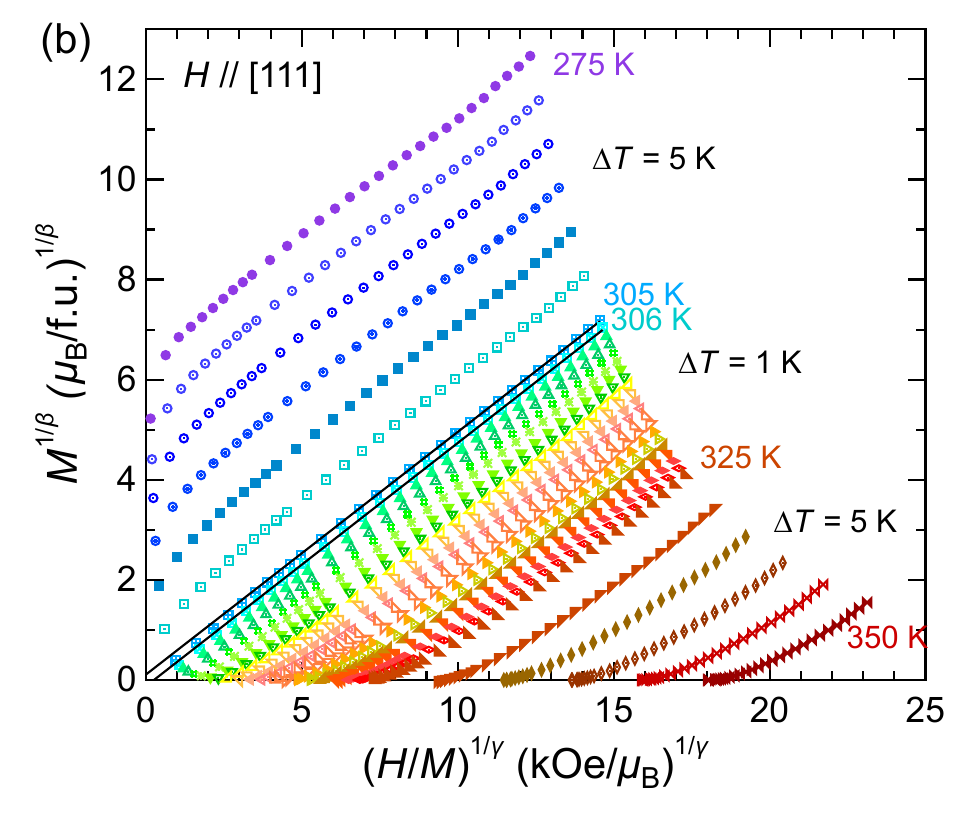}
    \end{minipage}\\
    \begin{minipage}[h]{0.48\linewidth}
      \centering
      \includegraphics[keepaspectratio, width=\columnwidth]{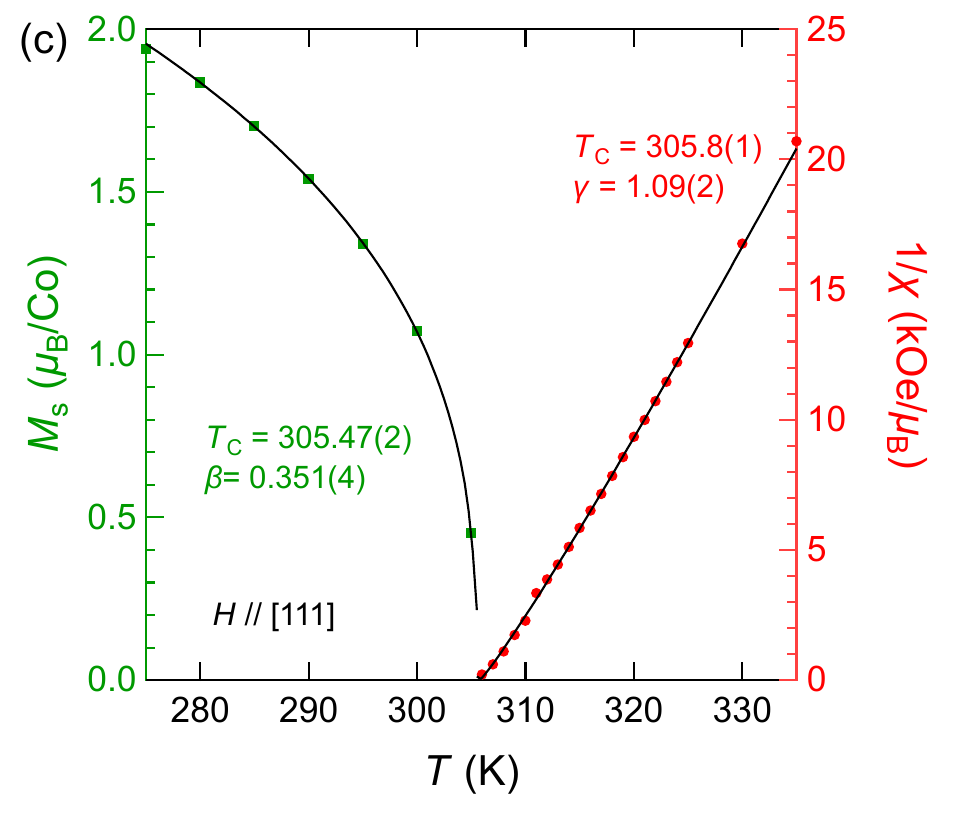}
    \end{minipage}&
    \begin{minipage}[h]{0.48\linewidth}
      \centering
      \includegraphics[keepaspectratio, width=\columnwidth]{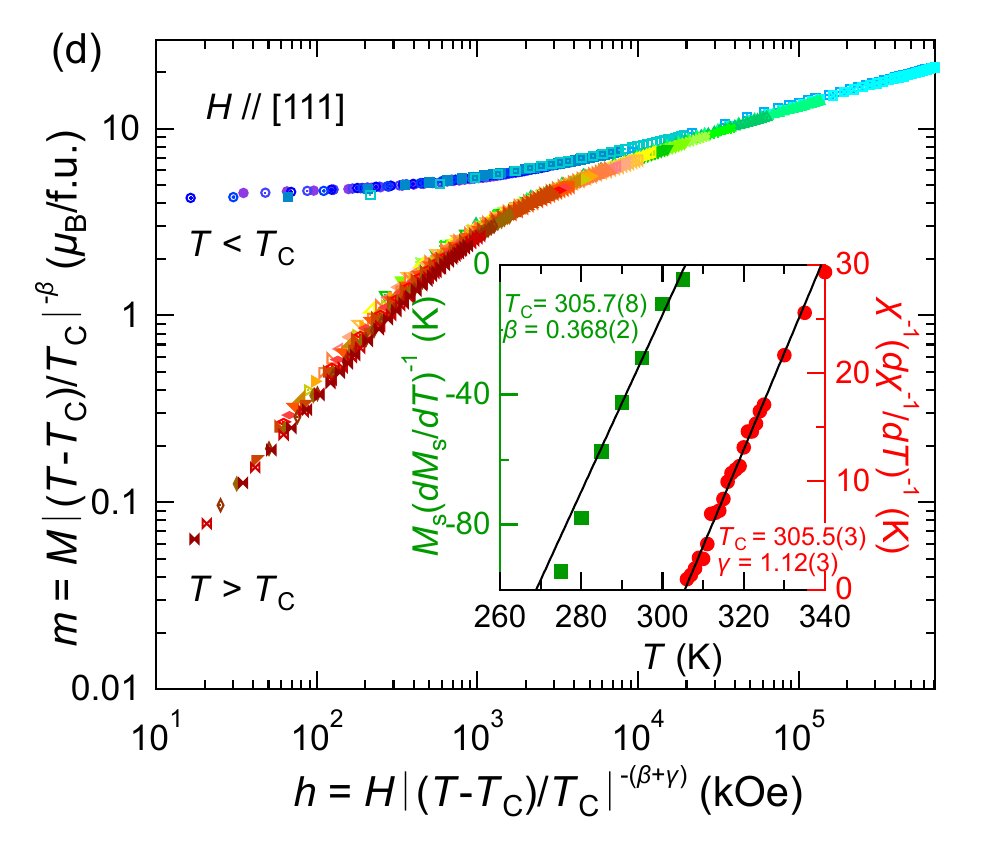}
    \end{minipage}
  \end{tabular}
    \caption{(a) The standard Arrott plots. The inset shows the isothermal magnetization at $T=306$\ K and the fitting results of $M=H^{\delta}$ with $\delta = 4.08$. (b) The final version of the generalized Arrott plots. (c) The temperature dependence of the spontaneous magnetization and inverse susceptibility obtained from the generalized Arrott plot. The lines represent the fit to the power function. (d) Scaling plots with $\beta$ and $\gamma$ determined by the KF method on both logarithmic scales. The inset shows the KF plots.}
    \label{Arrott}
  \end{figure*}
To precisely determine the Curie temperature and to study the critical behavior of {\LIC}, a standard Arrott plot ($M^2$ vs. $H/M$) was performed using field-dependent magnetization data measured at $T=275$--350 K [Fig.\ \ref{Arrott}(a)]. The Arrott plots have positive slope, in accordance with the nature of the second order transition.
Based on the mean-field approximation, the Arrott plot at $\TC$ should be straight and pass through the origin. As can be seen in Fig.\ \ref{Arrott}(a), it is not linear at low field, suggesting that the mean-field theory is too simple to explain the magnetism of {\LIC}. This deviation may be ascribed to the neglected electron correlations and spin fluctuations, which are significant in itinerant ferromagnets.
To better describe the magnetization process, a generalized Arrott plot (GAP) is known to be effective. A generalized Arrott plot is derived from the Arrott-Noaks equation of state \cite{AN},
\begin{equation}
  \left ( \frac{H}{M} \right )^{\frac{1}{\gamma}}=  a\epsilon + BM^{\frac{1}{\beta}},  
\end{equation}
where $\beta$ and $\gamma$ are the critical exponents, and the reduced temperature $\epsilon$ is given by $\epsilon = (T-\TC)/\TC$. 
Using $\beta$ and $\gamma$, the spontaneous magnetization $\Mspon$ and the inverse initial susceptibility $\invchi$ are described by the power laws \cite{KF},
\begin{align}
  &\Mspon (T) = M_0 (-\epsilon)^\beta , \quad T < \TC\ (\epsilon < 0), \label{mspon}\\
  &\invchi (T) =  \left (\frac{H}{M_0} \right )\epsilon^\beta , \quad T > \TC\ (\epsilon > 0), \label{invchi}
\end{align}
where $M_0$ and $H/M_0$ are the critical amplitudes. $\Mspon (T)$ and $\invchi (T)$ are estimated from the X- and Y-intercepts of the extrapolated lines in the high field region of the Arrott plots, respectively.
First, we obtained $\beta = 0.54$ and $\gamma = 0.98$ from $\Mspon (T)$ and $\invchi (T)$ estimated from the standard Arrott plots using the Eqs.\ (\ref{mspon}) and (\ref{invchi}), and used them to construct the modified Arrott plots. Again, we estimated $\Mspon (T)$ and $\invchi (T)$ from the established modified Arrott plots to obtain the critical exponents. This procedure was iterated until straight lines at $T \simeq \TC$ were obtained throughout the field range with unchanged critical exponents. 
Figure\ \ref{Arrott}(b) shows the final Arrott plots $M^{1/\beta}$ vs. $(H/M)^{1/\gamma}$. A good linearity can be seen near $\TC$, indicating that the critical exponents have been properly estimated ($\beta = 0.351(3), \TC = 305.47(3)$\ K and $\gamma= 1.09(2), \TC = 305.8(1)$\ K).
\\
\indent An alternative method for estimating the values of $\beta, \gamma$ and $\TC$ is the Kouvel--Fisher (KF) method \cite{KF}. Here, Eqs.\ (\ref{mspon}) and (\ref{invchi}) are reformulated as
\begin{align}
  &\Mspon (T) [d\Mspon (T)/dT]^{-1}= (T-\TC)\beta^{-1} , \quad T < \TC\ (\epsilon < 0), \label{kfmspon}\\
  &\invchi (T) [d\invchi (T)/dT]^{-1} =  (T-\TC)\gamma^{-1} , \quad T > \TC\ (\epsilon > 0). \label{kfinvchi}
\end{align}
$\beta$ and $\gamma$ can be calculated from the slope of $1/\beta$ and $1/\gamma$, and $\TC$ from the X-intercept in the plots of $\Mspon (T) [d\Mspon (T)/dT]^{-1}$ vs. $T$ and $\invchi (T) [d\invchi (T)/dT]^{-1}$ vs. $T$, respectively [Fig.\ \ref{Arrott}(c)].
Linear fits to Eqs.\ (\ref{kfmspon}) and (\ref{kfinvchi}) yield $\beta = 0.37(2), \TC = 305.7(8)$\ K and $\gamma= 1.12(3), \TC = 305.5(3)$\ K, respectively, which are close to those estimated from power-function fits. The small variation in $\beta$ is probably due to the limited magnetization data for $T<\TC$.
\\
\indent Finally, the critical exponent $\delta$ can be calculated using the equation \cite{KF}
\begin{equation}
  M \propto H^{1/\delta}.
\end{equation}
The fit yields $\delta = 4.08(3)$ [inset of Fig.\ \ref{Arrott}(a)], which is close to that estimated by the Widom relation $\delta = 1+\frac{\gamma}{\beta}$ \cite{Widom} (4.11(6) from the generalized Arrott plot and 4.03(22) from the KF method).
Based on the scaling hypothesis of the second order phase transition \cite{scaling}, the asymptotic equation of state near the transition temperature, the scaling form of the temperature-dependent magnetization is given by
\begin{equation}
  m=f_{\pm}(h),
\end{equation} 
where
\begin{equation}
  m=\abs{\epsilon}^{-\beta}M(H,\epsilon)\quad {\rm and} \quad h =H \abs{\epsilon}^{-(\beta + \gamma)}.
\end{equation}
$f_+(T<\TC)$ and $f_-(T>\TC)$ are the regular functions.
Figure \ref{Arrott}(d) shows the scaling plot $m$ vs. $h$ on both logarithmic scales obtained by using the critical exponents estimated by the KF method. The scaling curve is divided into two parts: the upper one is for $T<\TC$ and the lower one is for $T>\TC$, showing that the critical exponents $\beta$ and $\gamma$ and the Curie temperature $\TC$ have been properly estimated.
\\
\indent The obtained critical exponents satisfy none of the known universality classes, such as mean-field ($\beta = 0.5$, $\gamma = 1$, $\delta = 3$), three-dimensional (3D) Ising ($\beta = 0.326$, $\gamma = 1.238$, $\delta = 4.80$), 3D XY ($\beta = 0.345$, $\gamma = 1.316$, $\delta = 4.81$), and 3D Heisenberg model ($\beta = 0.367$, $\gamma = 1.388$, $\delta = 4.78$). 
While $\beta$ is close to the values in the three-dimensional universality class \cite{crt}, $\gamma$ is between the mean-field and three-dimensional universality classes. The discrepancy may be due to magnetic interactions beyond the nearest neighbor or short-range correlations.  
\\
\subsubsection{Magnetocaloric properties}
\begin{figure}[ht]
  \centering
  \includegraphics[keepaspectratio, width=1\columnwidth]{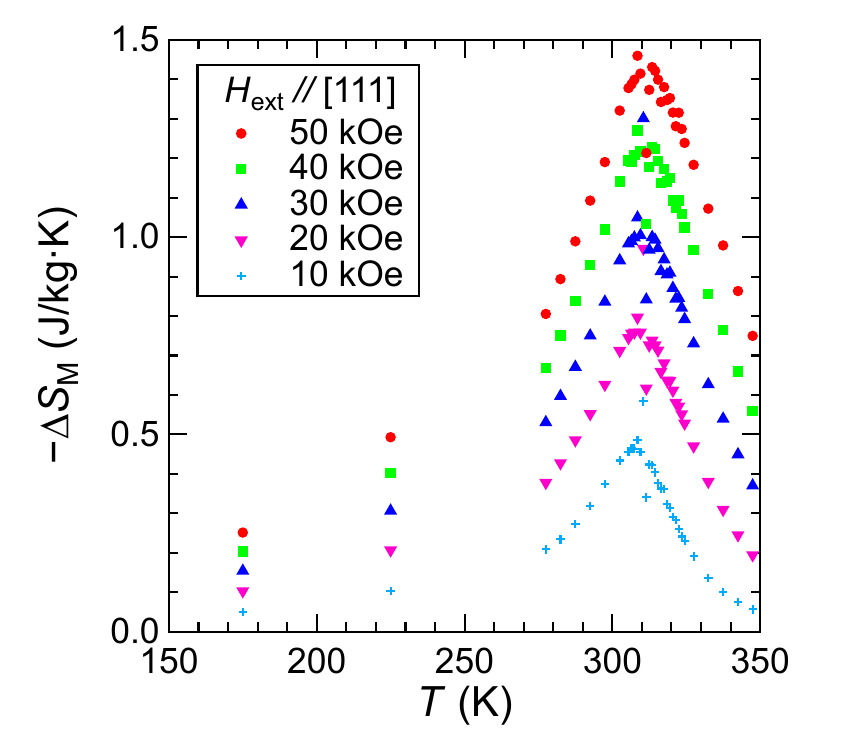}
  \caption{Temperature dependence of the magnetic entropy change $\dSM$ at selected fields applied in the [111] direction.}
  \label{MC}
\end{figure}
\indent Because of the ferromagnetic ordering of {\LIC} near room temperature, it is worth evaluating its magnetocaloric properties. The isothermal entropy change due to the application of a magnetic field $\dSM (T,M) = S(T,H)-S(T,0)$ is calculated by integrating the Maxwell relation as
\begin{equation}
\dSM = \int_{0}^{H}(\pdv*{M}{T})_{H'}dH'.
\end{equation}
In the real measurement of isothermal magnetization at a small step of field $\Delta H$, with a small temperature interval $T_2 - T_1$, $\dSM$ can be written by
\begin{equation}
  \dSM \left ( \frac{T_1+T_2}{2}, H \right) = \sum \left [ \frac{M(T_2, H)-M(T_1, H)}{T_2-T_1} \right ]\Delta H. 
\end{equation}  
Figure\ \ref{MC} shows the temperature dependence of $\dSM$ estimated from the field-dependent magnetization data at different temperatures around $\TC$ under the field applied in the [111] direction.
The maximum entropy change $-\dSMax$ related to the ferromagnetic ordering is indeed observed around the Curie temperature $T \simeq 310$\ K. The calculated values of $-\dSMax$ are 0.80(5) and 1.45(5) J/kg$\cdot$K for a field change of 0--2 T and 0--5 T, respectively. The maximum relative cooling power (RCP) is calculated as the product of the maximum entropy change $-\dSMax$ and the full-width at half-maximum of the $\dSM$ peak. The RCP values are estimated to be 30 and 105 J/kg, respectively. 
\\
\section{First-principle Calculations}
\begin{figure}[ht]
  \centering
  \includegraphics[keepaspectratio, width=1\columnwidth]{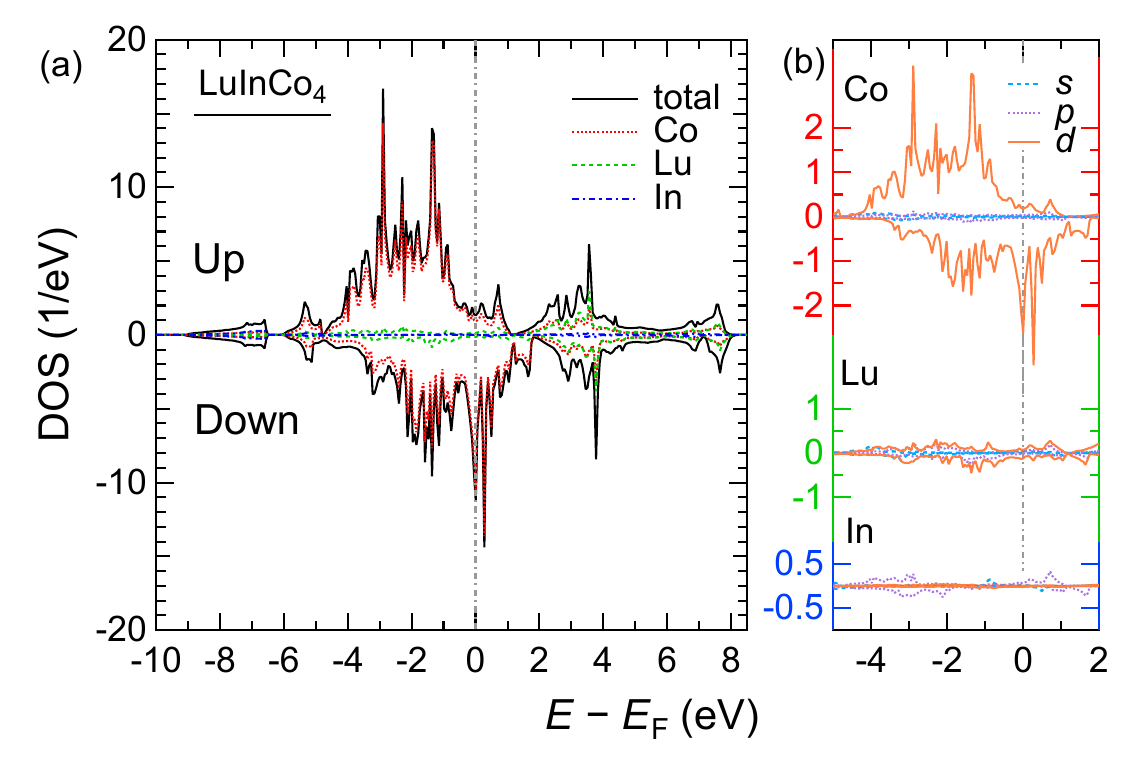}
  \caption{(a) Total and (b) site-decomposed DOS at each atomic site in the spin-polarized state of {\LIC} obtained from DFT calculations.}
  \label{DOS}
\end{figure}
\begin{figure*}[ht]
  \centering
  \includegraphics[keepaspectratio, width=1.75\columnwidth]{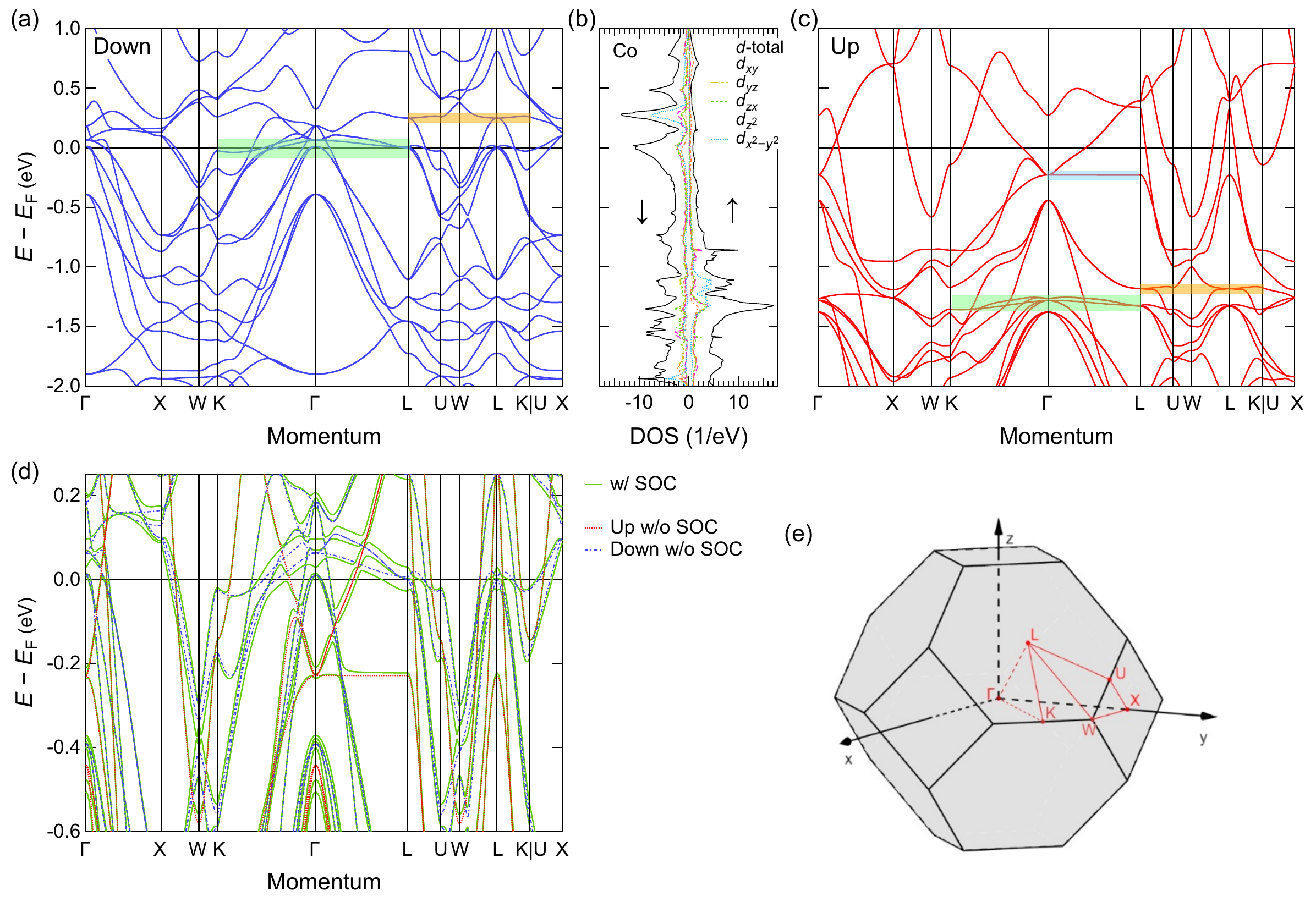}
  \caption{Band structure with down-spin (a) and up-spin (c) channels without SOC. (b) Orbital-decomposed DOS at the Co site in the spin-polarized state of {\LIC}. (d) Band structure with SOC when the [100] direction is the easy axis. (e) The first Brillouin zone of an FCC lattice.}
  \label{band}
\end{figure*} 
We have calculated the electronic structures in the spin-polarized (SP) and spin-unpolarized (NM) states to verify the ferromagnetic nature of {\LIC}. A comparison of the estimated total energies predicts that the SP state is more stable with a total energy 0.34 {eV/f.u.}\ lower than that of the NM state. 
The calculated equilibrium lattice parameters for the NM and SP states are $a_{\rm NM} = 6.9333 $\ {\AA} and $a_{\rm SP} = 6.9997$\ {\AA}, respectively. The latter is in better agreement with the experimental value (7.0387\ \AA) obtained from XRD measurements at room temperature. 
Figure\ \ref{DOS} shows the total DOS and the site-composed DOS of the SP state. The DOS near the Fermi energy ($\EF$) and the large band splitting are mainly due to the Co-3$d$ bands, which dominantly contribute to the ferromagnetic nature of {\LIC}. 
Similarly, the Lu-5$d$ DOS is found to split, which is caused by the hybridization between Co-3$d$ and Lu-5$d$ bands. 
The magnetic moment of each atom is calculated to be $\mu_{\rm Co}=1.290\ \uB, \mu_{\rm Lu}=-0.274$, and $\mu_{\rm In}=-0.042$. 
The total magnetic moment of 4.843 $\uB$ is slightly larger but comparable to the experimental value estimated by magnetization measurements (3.43 $\uB$).
Although the value of $\mu_{\rm Co}$ is not currently observed experimentally, it is close to the value obtained for {\YMC} by DFT calculations ($\mu_{\rm Co}=1.39$\ $\uB$ \cite{YMC}) and for {\LF} by neutron diffraction (ND) measurements ($\mu_{\rm Fe}=1.7$\ $\uB$ \cite{LuFeND, LuFeNDb}). 
We find that the Lu atom has a non-negligible magnetic moment, antiparallel to the Co moment, induced by the spin polarization of the Lu-5$d$ bands. It is well known that in such magnets containing 3$d$ transition metal elements and 5$d$ (4$d$) transition metal or rare earth elements, 3$d$--5$d$ (4$d$) hybridization plays an important role in the antiferromagnetic coupling between them \cite{ddhyba, ddhybb, ddhybc}.
When the 3$d$ bands are spin-polarized, the energy of the 3$d$-local DOS of the down spins becomes closer to the 5$d$-local DOS, which is at a higher energy level. Therefore, the 5$d$--3$d$ hybridization becomes stronger for down spin electrons, leading to the larger number of 5$d$ electrons with down spins below $\EF$.
Although not yet detected for {\LIC}, the presence of the magnetic moment $\mu_{\rm Lu}$ has been revealed by $^{175}$Lu-NMR measurement in {\LF} \cite{LuNMRLuFe}, and the ferrimagnetic structure between the Lu and Fe sublattices has been observed by spin-polarized ND \cite{LuFeND} and x-ray magnetic circular dichroism (XMCD) measurements \cite{XMCDa, XMCDb}.
\\
\indent Figures\ \ref{band}(a) and \ref{band}(c) show the down-spin and up-spin band structures, respectively. The orbital-resolved Co-3$d$ DOS is also shown in Fig.\ \ref{band}(b). 
In {\LIC}, the Co atoms form a pyrochlore lattice. As it contains three-dimensional stacking of the four equivalent \{111\}-kagome planes, the pyrochlore lattice can host three-dimensional flat bands throughout the Brillouin zone generated by destructive quantum interference of electron hopping \cite{topoFBpyro, FBPRB, CaNi}.
We found multiple narrow bands, as highlighted by the rectangular boxes in Figs.\ \ref{band}(a) and \ref{band}(c). 
Just above the $\EF$ for the down-spin channel, narrow bands lie along the K--$\rm{\Gamma}$--L path, forming a sharp peak of the Co-3$d$ DOS [Fig. \ref{band}(b)]. We can see from Fig. \ref{band}(b) that all $d$ orbitals contribute to the narrow bands. The corresponding bands for the up-spin channel appear at {$\sim$$-1.3$} eV, showing a splitting of {$\sim$1.3} eV by spin polarization. Interestingly, the bands are almost flat, suggesting localized states of the electrons reflecting the geometry of the pyrochlore lattice.  
Above these narrow bands, the DOS curve also has sharp peaks at {$\sim$0.25} and $-1.1$ eV for the down- and up-spin channels, respectively, which are attributed to narrow bands along the L--U and L--K paths. They are found to consist mainly of the $d_{x^2-y^2}$ orbital. 
Consequently, such highly degenerate electronic states lead to a strong Stoner instability, which induces the ferromagnetic correlations.
\\
\indent Since the heavy atom Lu is involved, we employed the spin-orbit coupling (SOC) in the calculations. We calculated the total energy for the collinear ferromagnetic structure with the spin oriented in the [100], [110] and [111] directions. 
It resulted in the lowest energy for the [111] easy axis case with the energy difference of $2.5 \times 10^{-6}$ {eV/f.u.}\ and $2.3 \times 10^{-5}$ {eV/f.u.}\ with the [100] and [110] cases, respectively. The lattice parameters and magnetic moments are the same as without SOC. Considering the energy tolerance of $10^{-7}$ {eV/f.u.}, it should be noted that the small energy difference is not accurate enough to discuss the spin orientation from DFT calculations, which may be affected by the initial parameters of the calculations. However, the detected small anisotropy is comparable to the experimental observation.
Figure\ \ref{band}(d) shows the band structure and partial DOS of the Co-$3d$ electrons calculated with SOC for the [100]\ easy axis. We observe the significant gap induced by SOC at the multiple band crossings and at the high-symmetry $\rm{\Gamma}$ and L points. The DOS is similar to that without SOC, including the occupancies of the orbitals. Note that similar results have been obtained for the [111] and [110]\ easy axes.
\\
\section{Discussion}
\begin{figure}[ht]
  \centering
  \includegraphics[keepaspectratio, width=1\columnwidth]{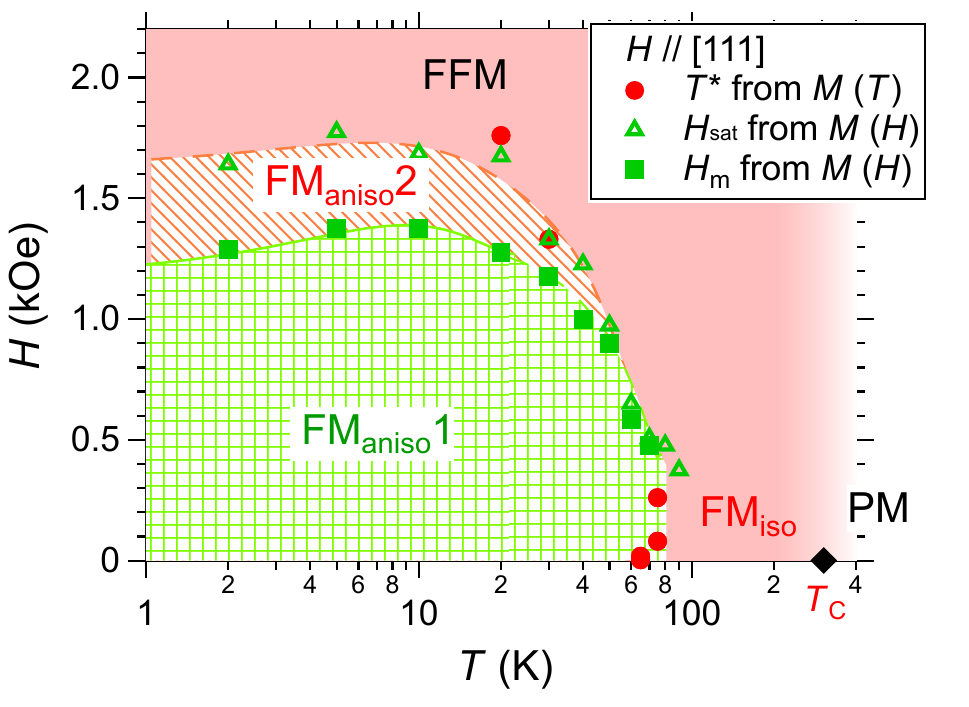}
  \caption{Inferred magnetic phase diagram of {\LIC} under the field applied in the [111] direction. $x$-axis (temperature) is shown on a logarithmic scale.}
  \label{PD}
\end{figure}
Figure\ \ref{PD} shows the magnetic phase diagram when the field is applied in the [111] direction as determined from the magnetization measurements. The temperature axis is plotted on a logarithmic scale. Remarkably, anisotropic ferromagnetic phases {\FMAo} and {\FMAt} appear below the temperature of $T \approx 100$\ K, in addition to the isotropic ferromagnetic phase (FM$_{\rm iso}$) just below the Curie temperature $\TC$.
The {\FMAo} phase is surrounded by the horizontal axis ($H=0$) and the metamagnetic-transition fields $\Hm$ (solid curve). 
The {\FMAt} phase is above the {\FMAo} phase, surrounded by the $\Hm$ curve and the fields where the magnetization is saturated, $H_{\rm sat}$, estimated from the second field derivative $d^2M/dH^2$ (dashed curve). This curve corresponds approximately to the field-dependent $\TCC$ in the temperature derivative $dM/dT$ [Fig.\ \ref{mtdmdt}(b)].
Thus, the {\FMAt} phase is considered to be one in which the magnetization is not saturated as a feature of the hard axis, and turns to the field-forced ferromagnetic (FFM) state.
Above $T \simeq 100$\ K, the magnetization is isotropic and quickly saturated, distinguishing the FM$_{\rm iso}$ phase from the {\FMAo} and {\FMAt} phases with the [111] hard axis.
\\
\indent Since the {\FMAo} phase has the large magnetization, three possible magnetic structures can be proposed. 
One is the collinear ferromagnetic structure, but with a smaller ordered moment than in the FFM phase. Due to the large 3$d$ down-spin DOS peak near the $\EF$ [Fig.\ \ref{DOS}], further instability and spin polarization can be induced by applying a magnetic field as a kind of itinerant metamagnetic transition \cite{itimagShimizu, itimagYamada} only in the [111] direction, accompanied by a discontinuous increase of the ordered moment. 
Similar magnetization jumps are observed in the ferromagnets {\ce{ThCo_5}} \cite{ThCo}, {\ce{YCo_3}} \cite{YCothr}, and {\ce{LuCo_3}} \cite{LuCothr}. This is reasonable because the down-spin bands lie along the $\Gamma (0,0,0)$--L($\frac{1}{2}$,$\frac{1}{2}$,$\frac{1}{2}$) direction [Fig.\ \ref{band}(a)].
Another is a weakly coupled ferrimagnetic structure between the Lu and Co sublattices. Although DFT calculations suggest that the ferrimagnetism is stable in the ground state, due to the weak magnetocrystalline anisotropy with $K_1=0.25$\ erg/cm$^3$, the Lu spins can be easily rotated toward the applied field direction and undergo a spin-flop transition when the field is applied in the [111] hard magnetization direction. 
From the magnetization curve at $T=5$\ K, the increase in magnetization at the metamagnetic transition could be roughly calculated to be {$\sim$0.2}\ $\uB$/f.u., which is of the same order of magnitude as $\mu_{\rm Lu}$ predicted by DFT calculations.
The other is the noncollinear structure of the Co sublattice, where the magnetic moments are practically ferromagnetic but have a small antiferromagnetic component. At the applied field of $\Hm$, the spins rotate towards the collinear ferromagnetic configuration (FFM). 
Here, we assume the situation with the nearest neighbor ferromagnetic coupling and the uniaxial magnetocrystalline anisotropy along the $\langle 111 \rangle$ directions. This is because the Co site forming the pyrochlore lattice is crystallographically unique with the locally trigonal symmetry axis along $\langle 111 \rangle$. Therefore, when the anisotropy is dominant, the moment at each Co site is constrained along four different symmetry directions, [\={1}11], [1\={1}1], [11\={1}], and [111], that is the spin-ice state \cite{spinice}. In this state, the [100], [110], and [111] directions are known to become the easy, hard, and hardest magnetization directions, respectively, with the different saturation magnetization ($4/\sqrt{3}\mu_{\rm Co}$, $2\sqrt{2/3}\mu_{\rm Co}$, and $2\mu_{\rm Co}$/f.u., respectively), and the metamagnetic transition occurs along the [111] direction \cite{icemeta, icemetab}. 
Except for the uniform saturation magnetization in all measured directions, the observed magnetization curves of {\LIC} at low temperature are similar to the feature of the spin-ice state. This suggests that the {\FMAo} phase may have unconventional spin-ice correlations as in ice-like splayed ferromagnetism \cite{splTi, splSn, splmag}, possibly due to the itinerant nature of the Co-3$d$ electrons and the insufficient uniaxial anisotropy. To gain a deeper insight, it is desirable to study the magnetization evolution at the lower temperature.\\

\section{Conclusion}
\indent We have successfully synthesized single crystals of the site-ordered cubic (C15b) Laves phase compound {\LIC} with a Co-pyrochlore sublattice. Magnetization measurements revealed that {\LIC} is a ferromagnet with $\TC = 306$\ K and $\Msat = 3.43$\ {$\uB$/f.u.}\ at $T=5$\ K.
DFT calculations predicted that the Co and Lu sublattices are antiferromagnetically coupled in the ground state, verifying the experimentally observed ferromagnetic nature. At high temperature the magnetization shows isotropic behavior, while at low temperature it exhibits magnetocrystalline anisotropy with the easy magnetization axis in the [100] direction. Moreover, the magnetization shows unusual ferromagnetic behavior: a metamagnetic transition in the [111] direction below the temperature where the magnetization becomes anisotropic, and critical behavior deviating from known universality classes.\\
\indent The indicated magnetically ordered phases in the low-$T$ and $H$ region and the multiple flat bands near the Fermi level makes {\LIC} an interesting ferromagnet from the point of view of the new frustrated and topological pyrochlore metals.
To reveal the microscopic magnetism at low temperature is highly desirable, for example by the neutron diffraction and nuclear magnetic resonance measurements on a single crystal.

\begin{acknowledgments}
This work was supported by JSPS KAKENHI Grant No. JP24KJ1325. 
\end{acknowledgments}

\bibliography{ref}

\end{document}